\documentclass{INTERSPEECH2023}

\usepackage{multirow}
\usepackage{makecell}
\usepackage{color}
\usepackage[ruled,linesnumbered]{algorithm2e}
\interspeechcameraready

\title{LibriheavyMix: A 20,000-Hour Dataset for Single-Channel Reverberant \\ Multi-Talker Speech Separation, ASR and Speaker Diarization}
\name{Zengrui Jin$_{*}^{1,3}$, Yifan Yang$_*^{1}$, Mohan Shi$_*^{2}$, Wei Kang$^1$, Xiaoyu Yang$^1$, Zengwei Yao$^1$, Fangjun Kuang$^1$, \\  Liyong Guo$^1$, Lingwei Meng$^3$, Long Lin$^1$, Yong Xu$^2$, Shi-Xiong Zhang$^2$, Daniel Povey$^1$ \thanks{* Equal contribution was made between the three authors.}}
\address{
  $^1$Xiaomi Corp., Beijing, China; 
  $^2$Tencent AI Lab, Bellevue, USA; \\
  $^3$The Chinese University of Hong Kong, Hong Kong SAR, China}
\email{zengrui.jin@link.cuhk.edu.hk; yifanyeung@sjtu.edu.cn; shimohan@g.ucla.edu}

\begin{document}
\bstctlcite{IEEEexample:BSTcontrol}

\maketitle

\begin{abstract}
The evolving speech processing landscape is increasingly focused on complex scenarios like meetings or cocktail parties with multiple simultaneous speakers and far-field conditions. Existing methodologies for addressing these challenges fall into two categories: multi-channel and single-channel solutions. Single-channel approaches, notable for their generality and convenience, do not require specific information about microphone arrays.

This paper presents a large-scale far-field overlapping speech dataset, crafted to advance research in speech separation, recognition, and speaker diarization. This dataset is a critical resource for decoding ``Who said What and When'' in multi-talker, reverberant environments, a daunting challenge in the field. Additionally, we introduce a pipeline system encompassing speech separation, recognition, and diarization as a foundational benchmark. Evaluations on the WHAMR! dataset validate the broad applicability of the proposed data.

\end{abstract}

\section{Introduction}

Despite the rapid progress of automatic speech recognition (ASR) technologies targeting single-talker, near-field speech \cite{wang2020transformer, gulati20_interspeech, yao2023zipformer}, these regular methods and datasets \cite{panayotov2015librispeech, aishell_2017, kang2023libriheavy} cannot handle the scenario where multiple speakers are presented simultaneously.

Existing works on speech separation \cite{hershey2016deep, YuKT017, luo2019conv, mossformer2} and multi-talker ASR \cite{yu2017recognizing,chang2019mimo,ZhangCQW20,kanda20b_interspeech,kanda22_interspeech, meng2023sidecar, meng2023unified} have been conducted on simulated multi-talker overlapping speech datasets \cite{hershey2016deep, Wichern2019WHAM, cosentino2020librimix,Maciejewski2020WHAMR}. 
However, most of these datasets neither take reverberation in the far-field condition into consideration, nor deliver sufficient amount of data for the model to be generalized to other datasets \cite{kadiouglu2020empirical, kanda2021large}. 
In addition, most of these datasets are simple cases with only 1 speaker turns, which does not match the real-world conversational scenarios where multiple speaker turns are common.
Recently, some real-world recorded multi-talker overlapping speech datasets \cite{carletta2005ami, yu2022m2met} are proposed with far-field reverberation and multiple speaker turns presented. 
However, the amount of data delivered by these datasets is still not large enough due to the very high recording cost. 
Moreover, it is difficult to obtain clean separation targets from real-world recorded data, limiting their capability as training data for speech separation models.

In this work, we propose a 20,000-hour multi-talker overlapping speech dataset LibriheavyMix based on Libriheavy \cite{kang2023libriheavy}, which is a
large-scale ASR corpus with richer information including punctuation casing and text context.
We conduct preliminary experiments on speech separation and multi-talker ASR on the proposed dataset and present the corresponding baseline results. 
Compared with previous work, LibriheavyMix presents the following advantages:
{\bf (1)} The amount of data is much larger than the others, with 10,000 hours.
{\bf (2)} Reverberation is introduced to simulate real-world far-field scenarios.
{\bf (3)} Multiple speaker turns, which is consistent with the real-world conversational scenarios, can be further used for speaker diarization \cite{park2022review,FujitaKHNW19,MedennikovKPKKS20,HeRHDC021} and speaker-attributed ASR \cite{KandaYGWMCY21, YuDZL022, shi23d_interspeech, ShiZDYCZD23}.
{\bf (4)} Punctuation, casing and text context are inherent in transcripts, which can be further combined with the research of punctuation and semantic information \cite{BijwadiaCLSZH22, shi23c_interspeech}.

\begin{table}[t]
	\caption{Statistics of simulated speech separation datasets. Note that the \# Hours listed for the training sets of the LibriheavyMix dataset is determined by summing the durations of all mixtures involving 1-4 speakers in total. }
	\label{tab:datasets}
	\centering
	\setlength\tabcolsep{2pt}
	\vspace{-8pt}
	\resizebox{\linewidth}{!}{
	\begin{tabular}{c|c|c|c|c|c|c}
        \toprule
		\toprule
        \multirow{2}{*}{Dataset} & \multirow{2}{*}{\makecell[c]{wsj0-mix \\ \cite{hershey2016deep}}} & \multirow{2}{*}{\makecell[c]{WHAM! \\ \cite{Wichern2019WHAM}}} & \multirow{2}{*}{\makecell[c]{Libri2Mix \\ \cite{cosentino2020librimix}}} & \multirow{2}{*}{\makecell[c]{Libri3Mix \\\cite{cosentino2020librimix}}} & \multirow{2}{*}{\makecell[c]{WHAMR! \\ \cite{Maciejewski2020WHAMR}}} & \multirow{2}{*}{\makecell[c]{\bf LibriheavyMix \\ (Ours)}} \\
        &&&&&& \\
        \midrule
        Reverberant & - & - & - & - & \textbf{\checkmark} & \textbf{\checkmark} \\
        \midrule
        Multi-Turn & - & - & - & - & - & \textbf{\checkmark} \\
        \midrule
        \multirow{4}{*}{Split} & train (30h) & train (30h)  & train-360 (212h) & train-360 (146h) & train (30h) & train-{\it small} (240h) \\
         & dev (8h) & dev (8h) & train-100 (58h) & train-100 (40h) & dev (8h) & train-{\it medium} (2,000h) \\
         & test (5h) & test (5h) & dev (11h) & dev (11h) & test (5h) & train-{\it large} (\textbf{18,000h}) \\
         & & & test (11h) & test (11h) & & \\
        % \midrule
        % \multirow{3}{*}{\# Hours} & 30 & 30 & 212 & 146 & 30 & 250 \\
        %  & 8 & 8 & 58 & 40 & 8 & 900 \\
        %  & 5 & 5 & 11 & 11 & 5 & 9000 \\
        %  & & & 11 & 11 & & \\
        \bottomrule
		\bottomrule
	\end{tabular}
	}
    \vspace{-20pt}
\end{table}

The rest of this paper is organized as follows. 
Section 2 presents the methods of data simulation. 
Section 3 shows the baseline systems of speech separation and multi-talker ASR. 
Section 4 shows experiments and results of baseline systems. 
Finally, Section 5 concludes this work.

% accurate separation and recognition of far-field overlapped speech remains a challenging task due to mixed signals from overlapping speakers and room reverberation. 
% These create a large mismatch between the reverberant mixture and dry clean speech. 

\section{Data Simulation}

\textbf{Simulation of Overlapped Speech: }
As described in Algorithm \ref{algo:simu}, $D_{={\rm spk}}$, $D_{\not={\rm spk}}$ and $D_{\rm ovlp}$ stand for the distribution of ``duration of pause between the same speaker'', ``duration of pause between two different speakers'' and ``duration of overlapping'' respectively. 
The statistics of these distributions are derived from the target sessions provided, and the duration sampled from the distribution is utilized to blend the source utterances.
Such a strategy is adopted as it has been successfully applied to improve end-to-end neural diarization \cite{landini22_interspeech}.
% TODO: add url for lhotse
%\footnote{ Implementation can be found in the Lhotse toolkit \cite{Zelasko_Lhotse_a_speech_2021}.}

\begin{algorithm}[t]
	\caption{Simulation of a session of $K$ speakers.} 
	\label{algo:simu} 
	\SetAlgoLined
    \KwData{$\mathcal{U}, D_{={\rm spk}}, D_{\not={\rm spk}}, D_{\rm ovlp}, P_{\rm ovlp}$}
    \KwResult{$\mathcal{M}$} 
	
		$\mathcal{M} \leftarrow \varnothing$
		
		$\mathcal{U} \leftarrow {\rm shuffle}(\mathcal{U}_{s_1}, \ldots, \mathcal{U}_{s_k})$
			
		${\rm offset} \leftarrow 0$, ${\rm num\_spk} \leftarrow 0$
			
		\For{$i \leftarrow 1$ \KwTo ${\rm range}(|\mathcal{U}|)$} {
				\eIf { $\mathcal{U}_i$.{\rm spk} == $\mathcal{U}_{i-1}$.{\rm spk} }{
					${\rm ot} \leftarrow {\rm sample}(D_{={\rm spk}})$
				} {
                    ${\rm num\_spk} += 1$ \\
					\lIf { \it Bernoulli($P_{\rm ovlp} > 0.5$) } {${\rm ot} \leftarrow - {\rm sample}(D_{\rm ovlp})$}
					\lElse {${\rm ot} \leftarrow {\rm sample}(D_{\not = {\rm spk}})$}
				}
				
				${\rm offset} \leftarrow {\rm offset} + {\rm ot} $
				
				$\mathcal{M} \leftarrow \mathcal{M} \cup \{U_i, {\rm offset}\}$ 

                    \lIf{${\rm num\_spk} == K$} {break}
		}
\end{algorithm}

Given the distribution of the target session on ``pause between the same speaker'' ($D_{={\rm spk}}$), ``pause between different speakers'' ($D_{\not={\rm spk}}$), ``duration of overlapping'' ($D_{\rm ovlp}$) and ``probability of overlapping'' ($P_{\rm ovlp}$),
a mixture $\mathcal{M}$ of $k$ speakers with a maximum duration of $T$ is simulated by first sampling source utterances from the provided samples $\mathcal{U} = \{\mathcal{U}_{s_1}, \ldots,  \mathcal{U}_{s_k}\}$ containing utterances from $\mathcal{S}$ speakers, $k$ denotes the index for distinct speakers.
The starting time of each of the selected utterances is sampled based on the speaker and the provided distributions $D_{={\rm spk}}$, $D_{\not ={\rm spk}}$ and $D_{\rm ovlp}$ as described in Algorithm \ref{algo:simu}.
An SNR value randomly selected within $[-5, 5]$ is assigned to utterances of each speaker before the segments are zero-padded and overlapped to form the {\it anechoic} single channel training samples. 

\noindent
\textbf{Simulation of Reverberation: }
%Reverberation was also introduced using FAST-RIR \cite{fast-rir} \footnote{Implementation can be found in \url{https://github.com/anton-jeran/FAST-RIR/}.}, which provides GPU accelerated GAN-based model to generate room impulse responses and convolved with dry clean source utterances to extend the simulated data to a more challenging reverberant scenario.
% TODO: add the url for fast-rir
Reverberation was also introduced using FAST-RIR \cite{fast-rir}, which provides GPU accelerated GAN-based model to generate room impulse responses and convolved with dry clean source utterances to extend the simulated data to a more challenging reverberant scenario.

Given a session $\mathcal{U} = \{\mathcal{U}_1, \ldots, \mathcal{U}_k\}$ composed of source segments from $k$ speakers, the reverberation time ($T_{60}$), room dimension ($RD$) and listener position ($LP$) are identical for all sources in the same session to form a consistent acoustic environment.
Meanwhile, the source position ($SP$) for each speaker is slightly perturbed within the range of previously set $RD$ to model the variation of positions of each source speaker.
$D_{RD}$ and $D_{T_{60}}$ indicate the distribution of room dimension and reverberation time. 
As described in Algorithm \ref{algo:rir}, each source is convolved with the room impulse response of an identical acoustic environment, but with various locations generated by FAST-RIR. 
This results in a reverberant session $\mathcal{U}'$.
The {\it reverberant} mixture $\mathcal{M}'$ can be derived from $\mathcal{U}'$ and offsets of the original $\mathcal{M}$.

\noindent
\textbf{Libriheavy and LibriheavyMix:}
To simulate the LibriheavyMix dataset with a realistic distribution, corresponding statistics are obtained from the AliMeeting \cite{yu2022m2met} dataset, which is a publicly available conference scenario dataset with human-annotated segmentation.
Source utterances are from the Libriheavy \cite{kang2023libriheavy} dataset, which is an ASR corpus for large-scale supervised training consisting of 50,000 hours of data.
The Libriheavy dataset provides richer information for system construction such as punctuation, casing, and text context, which are also provided along with the speaker identity and corresponding timestamps for further investigation.
During simulation, mixtures are generated by randomly selecting utterances for different speakers, each speaker is assigned with no longer than 15 seconds of the source utterances.
Utterances with a duration longer than 15 seconds are first aligned using wav2vec2.0 \cite{baevski2020wav2vec} to obtain boundaries of sub-utterances for mixture simulation. 
Unlike the wsj0-mix \cite{hershey2016deep}, each utterance from the Libriheavy training set is used only once during the process of simulation, creating enough diversity in the simulated training data.
The simulated dataset is provided with a {\it max} and {\it min} version, shorter sources in the {\it max} version are padded to the longest one, while mixtures in the {\it min} version were truncated to align with the source with shortest duration. 
This results in approximately 100 hours, 900 hours, and 9,000 hours of data in the {\it max} version of the small, medium, and large training set, against 45 hours of the wsj-mix dataset and 456 hours of the LibriMix dataset. 
Each training set of LibriheavyMix uniformly includes conversations involving 1-4 speakers, noted as {\it small\{1-4\}spk}, {\it medium\{1-4\}spk} and {\it large\{1-4\}spk}, respectively. 
For the dev and test sets, mixtures containing 2 to 4 speakers are derived from the dev, test-clean and test-other sets of the original Libriheavy corpus, noted as {\it dev\{2-4\}spk}, {\it test-clean\{2-4\}spk} and {\it test-other\{2-4\}spk}, respectively. 
The variety of speakers is much wider in LibriheavyMix's training set with around 6,000 distinct speakers in the {\it large} training set against 1,000 speakers in LibriMix and 100 speakers in wsj0-mix.

\begin{algorithm}[t]
	\caption{Simulation of the reverberant session.}
	\label{algo:rir}
	\SetAlgoLined
    \KwData{$\mathcal{U} = \{\mathcal{U}_1, \ldots, \mathcal{U}_k\}$, $D_{RD}$,  $D_{T_{60}}$}
    \KwResult{$\mathcal{U}'  = \{\mathcal{U}_1', \ldots, \mathcal{U}_k'\}$} 
    
    $RD_X, RD_Y, RD_Z \gets {\rm sample}(D_{RD})$
    
    $LP_X, LP_Y, LP_Z \gets {\rm sample}(D_{RD})$
    
    $T_{60} \gets {\rm sample}(D_{T_{60}})$
    
    $\mathcal{X}' \gets \varnothing$

    \For{$i \gets 1$ \KwTo $k$} {
	    $SP_X, SP_Y, SP_Z \gets {\rm sample}(D_{RD_{X, Y, Z}})$

	    $\mathcal{U}'_i \gets {\rm FAST\text{-}RIR}(\mathcal{U}_i; RD_{X, Y, Z}; LP_{X, Y, Z}; SP_{X, Y, Z}; T_{60})$
	    
	    $\mathcal{U}' \gets \mathcal{U}' \cup \{ \mathcal{U}'_i \}$
    }
\end{algorithm}

\section{Baseline Systems}

\begin{figure}[t]
    \centering
    \includegraphics[width=\linewidth]{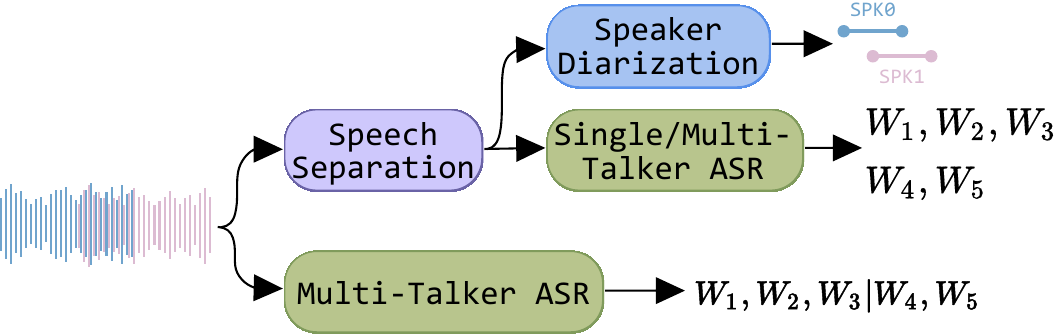}
    \caption{Pipeline system of Separation, Diarization and ASR. }
    \label{fig:sys}
    \vspace{-15pt}
\end{figure}

The pipeline system constructed involves a multi-talker ASR system, speech separation model and a diarization system as illustrated in Fig. \ref{fig:sys}.

\subsection{Baseline for Multi-Talker ASR}

For the multi-talker ASR baseline system, serialized output training (SOT) \cite{kanda20b_interspeech} based Conformer \cite{gulati20_interspeech} Attention Encoder Decoder (AED) model is employed.
Given input $\boldsymbol{X} = \{x_1, \cdots , x_T \}$, a single-speaker AED model produces the output sequence $\boldsymbol{Y} = \{y_1, \cdots, y_{N}\}$ as follows. 
Firstly, the encoder converts the input sequence $\boldsymbol{X}$ into a sequence embeddings by
\begin{equation}
	\boldsymbol{H}^{enc}=\left\{h_1^{enc}, \cdots, h_T^{enc}\right\}=\operatorname{Encoder}(\boldsymbol{X})
\end{equation}
Then, given the previous output $y_{[1:n-1]}$ and the encoder embeddings $\boldsymbol{H}^{enc}$, the output $y_n$ is estimated by the autoregressive Transformer-based decoder.
\begin{equation}
	y_n=\operatorname{Decoder}\left(y_{[1:n-1]}, \boldsymbol{H}^{enc}\right)
\end{equation}

% Then, given the previous attention weight $\alpha_{n-1}$ and the encoder embeddings $\boldsymbol{H}^{enc}$, the attention weight $\alpha_n$ and context vector $c_n$ of the $n$-th decoder step is given by
% \begin{equation}
% 	c_n, \alpha_n=\operatorname{Attention}\left(q_n, \alpha_{n-1}, \boldsymbol{H}^{enc}\right)
% \end{equation}
% Finally, the output $y_n$ is estimated given the context vector $c_n$ and the decoder state vector $q_n$ as follows:
% \begin{equation}
% q_n=\operatorname{Decoder}\left(y_{n-1}, c_{n-1}, q_{n-1}\right)
% \end{equation}
% \begin{equation}
% 	y_n=\operatorname{DecoderOut}\left(c_n, q_n\right)
% \end{equation}
% where $\operatorname{DecoderOut}$ can be implemented with an affine transform with a softmax output layer.

To incorporate the SOT paradigm, a special symbol $\langle sc \rangle$ is inserted in the concatenation of multiple references to represent ``speaker change'' between each turn. 
Given a two-speaker conversation with speaker A and B, the reference word sequence will be given as $\boldsymbol{W}=\{\boldsymbol{w}_A^1, \cdots, \boldsymbol{w}_A^n, \langle sc \rangle, \boldsymbol{w}_B^1, \cdots, \boldsymbol{w}_B^m, \langle sc \rangle, \boldsymbol{w}_A^1, \cdots, \boldsymbol{w}_A^o \}$,  where $n$, $m$ and $o$ represent number of tokens in the transcript of each utterance.
Reference labels in the resulting concatenation $\boldsymbol{W}$ are sorted by their start times in a first-in, first-out fashion. 
In this way, the AED model learns to identify the turning point in a given utterance of multiple speakers marked by the special symbol $\langle sc \rangle$, thereby separating the transcript.

\subsection{Baseline for Speech Separation}

\begin{table}[t]
  \caption{Word error rate (\%) of the AED model pre-trained on  FAST-RIR augmented LibriSpeech 100-hour data on LibriheavyMix test sets of 2 to 4 speakers.}
  \label{tab:pretrained}
  \centering
  \vspace{-8pt}
  \renewcommand\tabcolsep{3.0pt}
  \resizebox{\linewidth}{!}{
  \begin{tabular}{c|c|c|c|c|c|c|c|c|c|c|c}
  	\toprule
  	\toprule
  	\multirow{3}{*}{Sys.} & \multirow{3}{*}{\makecell[c]{Training \\ Set}} & \multirow{3}{*}{\makecell[c]{Test Set \\ \# spkr}} & \multicolumn{3}{c|}{dry clean} & \multicolumn{3}{c|}{reverb clean} & \multicolumn{3}{c}{reverb mixture} \\
	  &  &  & \multirow{2}{*}{dev} & \multirow{2}{*}{\makecell[c]{test-\\clean}} & \multirow{2}{*}{\makecell[c]{test-\\other}} & \multirow{2}{*}{dev} & \multirow{2}{*}{\makecell[c]{test-\\clean}} & \multirow{2}{*}{\makecell[c]{test-\\other}} & \multirow{2}{*}{dev} & \multirow{2}{*}{\makecell[c]{test-\\clean}} & \multirow{2}{*}{\makecell[c]{test-\\other}} \\
  	  &  &  &  &  &  &  &  &  &  &\\
	\midrule
	\multirow{3}{*}{1} & \multirow{3}{*}{\makecell[c]{LS100\\w. RIR}} & 2 & 35.6 & 34.3 & 39.3 & 35.2 & 34.0 & 38.2 & 106.2 & 114.0 & 106.7 \\
  	&  & 3 & 35.3 & 31.1 & 38.7 & 33.8 & 29.6 & 37.8 & 121.6 & 122.3 & 121.6 \\
  	&  & 4 & 34.3 & 29.3 & 39.0 & 33.8 & 27.4 & 37.5 & 130.9 & 139.0 & 137.7 \\
  	\bottomrule
  	\bottomrule
  \end{tabular}
  }
  \vspace{-15pt}
\end{table}

The Conv-TasNet \cite{luo2019conv} is selected as the baseline system for speech separation task. 
The model is a fully convolutional model specifically designed to separate individual speakers from a given mixed time-domain segment $\mathbf{x} \in \mathbb{R}^{1\times L}$, where $L$ represents the number of samples of the given mixture. 
The network involves three stages: encoder, separation, and decoder. 
Encoder maps the segment to a high-dimensional representation $\boldsymbol{H}^{enc}$ using 
\begin{equation}
	\boldsymbol{H}^{enc} = \mathcal{H}(\mathbf{x}\mathbf{U})
\end{equation}
where $\mathbf{U} \in \mathbb{R}^{N\times L}$ represents 1-D convolution operations with $N$ kernels, each of length $L$.
This operation can be represented as a matrix multiplication.
$\mathcal{H}(\cdot)$ is a rectified linear unit (ReLU) to ensure the non-negativity of $\boldsymbol{H}^{enc}$.
The separation stage involves a series of 1-D convolution blocks of different dilation factors to estimate the masks for each of the target sources based on the encoder output.
Estimated masks are multiplied with the encoded high-dimensional representation, a decoder further reconstructs the masked feature to waveforms using a 1-D transposed convolution operation as 
\begin{equation}
	\tilde{\mathbf{x}} = \tilde{\boldsymbol{H}}^{enc}\ \mathbf{V}
\end{equation}
where $\tilde{\boldsymbol{H}}^{enc}$ and $\tilde{\mathbf{x}}$ stand for the masked feature and reconstructed waveforms, $\mathbf{V} \in \mathbb{R}^{N \times L}$ represents $N$ kernels of the convolution operation, each with a dimension of $L$.

The model is trained end-to-end by minimizing the negative scale-invariant source-to-noise ratio (SI-SNR) loss $\mathcal{L}_{\text{SI-SNR}}$ given by
\begin{equation}
\mathcal{L}_{\text{SI-SNR}}=-10 \log_{10} \frac{||\mathbf{s}_{target}||^2}{||\mathbf{e}_{noise}||^2}
\end{equation}
in which $\mathbf{s}_{target}$, $\mathbf{e}_{noise}$ are obtained through $\mathbf{s}_{target}= \frac{\langle \hat{\mathbf{s}},\mathbf{s}\rangle \mathbf{s}}{\langle \mathbf{s},\mathbf{s}\rangle}$ and $\mathbf{e}_{noise} = \hat{\mathbf{s}} - 
\mathbf{s}_{target}$ given estimated sources $\hat{\mathbf{s}} \in \mathbb{R}^{1\times T}$ and original clean sources $\mathbf{s} \in \mathbb{R}^{1\times T}$. 
$\hat{\mathbf{s}}$ and $\mathbf{s}$ are normalized to zero-mean before loss calculation.
To address the source permutation problem, utterance-level permutation invariant training (uPIT) \cite{YuKT017} is incorporated during training.

%\subsection{SpeakerBeam Model}
%
%The SpeakerBeam \cite{vzmolikova2019speakerbeam} model adopts a similar temporal convolutional architecture and masking-based strategy as in the Conv-TasNet.
%Confronting the target speaker extraction issue, the SpeakerBeam model learns to separate the signal of a target speaker from the unwanted mixture using auxiliary information. 
%Unlike previous systems using i-Vector to represent speaker-related information, the SpeakerBeam model features an additional sequence summarizing network to directly obtain speaker information from a given adaptation utterance $\mathbf{x}^{aux}$ of the target speaker by
%\begin{equation}
%	\boldsymbol{H}^{aux} = \mathcal{J}(\mathbf{x}_{aux})
%\end{equation}
%where $\boldsymbol{H}^{aux} = \{h_1^{aux}, \cdots, h_T^{aux}\}$ is the high-dimensional representation of the adaptation utterance $\mathbf{x}^{aux}$ encoded by auxiliary neural network $\mathcal{J}$.
%$\boldsymbol{H}^{aux}$ is further averaged along the temporal axis to obtain the final speaker-related information $\lambda^{aux}$, which will be further incorporated with the mask estimation process to generate a mask for the indicated speaker.
%

\subsection{Baseline for Speaker Diarization}
Pre-trained pyannote.audio 3.1 system \footnote{\url{https://huggingface.co/pyannote/speaker-diarization-3.1/}} \cite{bredin23_interspeech, plaquet23_interspeech} is applied as the baseline system for speaker diarization experiments.
The system first utilizes a neural speaker segmentation model, incorporating a sliding window to obtain local speaker segmentation.
Local speaker embeddings are then extracted from each window, and classical agglomerative hierarchical clustering with centroid linkage is then applied to the extracted embeddings.
The final aggregating step produces the actual speaker diarization results on top of the clustered local speaker segmentation.

\section{Experiments and Results}

\subsection{Performance of Multi-Talker ASR Baseline}

\begin{table}[t]
  \caption{Performance of the Serialized Output Training (SOT) \cite{kanda20b_interspeech} models on dev/test sets of 2 to 4 speakers. Systems are initialized using the single channel ASR model in Table \ref{tab:pretrained} (Sys. 1) and trained on the small, medium and large in the ``Training Set'' column stand for the small\{1-4\}spk, medium\{1-4\}spk and large\{1-4\}spk training sets. cpWER represents the concatenated minimum-permutation word error rate \cite{watanabe2020chime}.}
  \label{tab:sot_aed_var_spkr}
  \centering
  \vspace{-8pt}
  \resizebox{\linewidth}{!}{
  \begin{tabular}{c|c|c|c|c|c|c|c|c}
  	\toprule
  	\toprule
  	\multirow{3}{*}{Sys.} & \multirow{3}{*}{\makecell[c]{Training \\ Set}} & \multirow{3}{*}{\makecell[c]{Test Set \\ \# spkr}} & \multicolumn{3}{c|}{cpWER (\%) $\downarrow$} & \multicolumn{3}{c}{Spkr. Counting Acc. (\%) $\uparrow$} \\
	  &  &  & \multirow{2}{*}{dev} & \multirow{2}{*}{\makecell[c]{test-\\clean}} & \multirow{2}{*}{\makecell[c]{test-\\other}} & \multirow{2}{*}{dev} & \multirow{2}{*}{\makecell[c]{test-\\clean}} & \multirow{2}{*}{\makecell[c]{test-\\other}} \\
	&&&&&&&& \\
	\midrule
  	\multirow{3}{*}{1} & \multirow{3}{*}{\makecell[c]{{\it small}}} & 2 & 57.8 & 59.5 & 61.0 & 45.19 & 42.56 & 46.84 \\
  	&  & 3 & 68.3 & 70.5 & 73.4 & 33.51 & 25.97 & 28.21 \\
  	&  & 4 & 76.0 & 76.3 & 79.5 & 25.22 & 25.04 & 26.85 \\
  	\midrule
  	\multirow{3}{*}{2} & \multirow{3}{*}{{\it medium}} & 2 & 27.2 & 25.7 & 27.5 & 54.12 & 52.89 & 56.55 \\
  	&  & 3 & 35.8 & 35.8 & 40.4 & 41.04 & 33.20 & 38.16 \\
  	&  & 4 & 52.3 & 48.3 & 54.5 & 22.02 & 20.66 & 21.39 \\
  	\midrule
  	\multirow{3}{*}{3} & \multirow{3}{*}{{\it large}} & 2 & 21.0 & 19.0 & 21.7 & 55.48 & 54.18 & 56.25 \\
  	&  & 3 & 28.8 & 27.7 & 31.7 & 41.48 & 37.84 & 41.01 \\
  	&  & 4 & 40.4 & 38.9 & 43.3 & 22.63 & 20.59 & 21.62 \\
  	\bottomrule
  	\bottomrule
  \end{tabular}
  }
  \vspace{-15pt}
\end{table}

The recipe for serialized output training (SOT) \cite{kanda20b_interspeech} is modified from the LibriMix recipe of the ESPnet \cite{watanabe2018espnet} toolkit.
Trained Conformer models are of 12 encoder blocks and 6 Transformer decoder blocks, with a total of 43 million parameters.
The dimension of feed forward layers in both encoder and decoder blocks is set to 2048 with 4 attention heads, each attention head has a dimension of 256, kernel size of all convolutional layers is set to 31\footnote{ More details can be found at \url{egs2/librimix/sot_asr1/conf/tuning/train_sot_asr_conformer.yaml} of the ESPnet toolkit \cite{watanabe2018espnet}.}.
To help convergence, systems trained were initialized using a Conformer model with an identical setup pre-trained on LibriSpeech \cite{panayotov2015librispeech} 100-hour training set augmented using FAST-RIR.
Performance of the pre-trained system is presented in Tab. \ref{tab:pretrained}, Sys. 1.
The training data includes all available mixtures, the transcript of which is obtained by concatenating the transcript of each source according to its starting time in a ``first-in, first-out'' fashion. 
SpecAugment \cite{Park2019} is incorporated for all systems.
Speed perturbation \cite{ko2015audio} is further applied except for the ones with {\it large\{1-4\}spk} training set involved. 
The evaluation metric for all results obtained from SOT systems is the concatenated minimum-permutation word error rate (cpWER) \cite{watanabe2020chime}. This metric is calculated by first concatenating all utterances of each speaker for both the reference and hypothesis files. Then, the permutation of speakers that yields the lowest word error rate when compared to the reference is picked.

Sys. 1-3 (Tab. \ref{tab:sot_aed_var_spkr}) were trained on the small, medium, and large training sets of the proposed LibriheavyMix corpus respectively. 
All training data containing 1 to 4 speakers were involved to evaluate the capability of SOT systems on generalizing to mixtures of various numbers of speakers. 
Results suggest that scaling up the amount of training data demonstrates a significant reduction on cpWER and speaker counting accuracy.
Sys. 2 consistently outperforms Sys. 1 on all test sets, while a similar trend can also be observed between Sys. 3 and 2 except for a slight performance degradation on speaker counting accuracy is obtained on the most challenging 4-speaker scenario. 

\subsection{Performance of Speech Separation Baseline}

\begin{table}[t]
  \caption{Performance of the Conv-TasNet \cite{luo2019conv} models on the LibriheavyMix and WHAMR! dataset. ``tt'' stands for the test set of WHAMR!. ``dev'', ``test-clean'' and ``test-other'' denote the dev2spk, test-clean2spk and test-other2spk of LibriheavyMix. }
  \label{tab:whamr_enh}
  \centering
  \vspace{-8pt}
  \renewcommand\tabcolsep{2.5pt}
  \resizebox{\linewidth}{!}{
  \begin{tabular}{c|c|c|c|c|c|c|c|c|c|c|c|c}
  	\toprule
  	\toprule
  	\multirow{3}{*}{Sys.} & \multicolumn{4}{c|}{Training Set} & \multicolumn{4}{c|}{SI-SDR $\uparrow$} & \multicolumn{4}{c}{$\Delta$ SI-SDR $\uparrow$}  \\
	  & \multicolumn{3}{c|}{LibriheavyMix ({\it 2spk})} & \multirow{2}{*}{WHAMR!} & \multirow{2}{*}{tt} & \multirow{2}{*}{dev} & \multirow{2}{*}{\makecell[c]{test-\\clean}} & \multirow{2}{*}{\makecell[c]{test-\\other}} & \multirow{2}{*}{tt} & \multirow{2}{*}{dev} & \multirow{2}{*}{\makecell[c]{test-\\clean}} & \multirow{2}{*}{\makecell[c]{test-\\other}} \\
  	  & small & medium & large &  &  &  &  &  &  &  &  \\
	\midrule
  	1 & \multicolumn{3}{c|}{-} & \checkmark & 9.36 & 1.36 & 1.99 & 1.51 & 9.36 & 1.95 & 2.04 & 1.65  \\
  	\midrule
  	2 & \checkmark &  &  & - & 5.11 & 6.41 & 7.83 & 6.14 & 5.11 & 7.08 & 7.88 & 6.28  \\
  	3 &  & \checkmark &  & - & 8.19 & 9.27 & 11.33 & 10.11 & 8.20 & 9.86 & 11.37 & 10.25  \\
  	4 &  &  & \checkmark & - & 9.23 & 10.70 & 12.94 & 11.54 & 9.23 & 11.55 & 12.90 & 11.52 \\
  	\midrule
  	5 & \checkmark &  &  & \checkmark & 10.02 & 7.24 & 9.19 & 7.53 & 10.02 & 7.83 & 9.24 & 7.67 \\
  	6 &  & \checkmark &  & \checkmark & 10.49 & 9.75 & 12.06 & 10.58 & 10.49 & 10.33 & 12.11 & 10.72 \\
   	7 &  &  & \checkmark & \checkmark & 10.33 & 10.66 & 12.81 & 11.35 & 10.34 & 11.24 & 12.87 & 11.49  \\
  	\bottomrule
  	\bottomrule
  \end{tabular}
  }
  \vspace{-8pt}
\end{table}

%The Conv-TasNet \cite{luo2019conv} model trained is of 8.98 million of parameters \footnote{ Implementation can be found at: \url{https://github.com/funcwj/conv-tasnet}}.
% TODO: add url for conv-tasnet
The Conv-TasNet \cite{luo2019conv} model trained has 8.98 million parameters.
The encoder contains 512 filters, the length of each filter is set to 40,  bottleneck dimension is set to 256.
The repeat number is set to 4, each repeat contains 8 convolutional blocks with kernel size set to 3 and number of channels set to 512. 
Global layer normalization and ReLU are adopted for normalization and non-linearity respectively. 
Training is done by minimizing the negative permutation-invariant, SI-SNR loss on 4-second segments.
All systems were trained with identical parameters.
Since the SI-SDR is not defined for silent sources, all results reported were trained on the 8kHz {\it min} version of the training sets and evaluated on the {\it max} version of test sets. 

The performance of the Conv-TasNet model on LibriheavyMix dataset is presented in Tab. \ref{tab:whamr_enh}, Sys. 2-4. 
All models were trained and evaluated on the 2-speaker sets of LibriheavyMix.
Results suggest that scaling up the training data demonstrates significant performance improvements on all test sets, as Sys. 4 trained on approx. 7,000-hour {\it large2spk} consistently outperforms Sys. 3 trained on approx. 500-hour {\it medium2spk} set.
A similar trend is also obtained on Sys. 3 and Sys. 2 trained on the approx. 70-hour {\it small2spk} set.

%\subsection{Performance of SpeakerBeam Model}
%
%The SpeakerBeam \cite{vzmolikova2019speakerbeam} model trained is of 13.5 million of parameters \footnote{Implementation can be found at: \url{https://github.com/BUTSpeechFIT/speakerbeam}}
%
%\input{tab/speakerbeam_2_mix}

%\input{tab/speakerbeam_var_mix}

\subsection{Generalization on the WHAMR! dataset}

\begin{table}[t]
  \caption{Performance of the SOT \cite{kanda20b_interspeech} models on the WHAMR! dataset. Other naming conventions follow the one in Table \ref{tab:whamr_enh}. Note that only performance on {\bf 2-speaker} test sets are presented in this table for simplicity.}
  \label{tab:whamr_asr}
  \centering
  \vspace{-8pt}
  \renewcommand\tabcolsep{3.0pt}
  \resizebox{\linewidth}{!}{
  \begin{tabular}{c|c|c|c|c|c|c|c|c|c|c|c|c}
  	\toprule
  	\toprule
  	\multirow{3}{*}{Sys.} & \multicolumn{4}{c|}{Training Set} & \multicolumn{4}{c|}{cpWER (\%) $\downarrow$} & \multicolumn{4}{c}{Spkr. Counting Acc. (\%) $\uparrow$}  \\
	  & \multicolumn{3}{c|}{LibriheavyMix} & \multirow{2}{*}{WHAMR!} & \multirow{2}{*}{tt} & \multirow{2}{*}{dev} & \multirow{2}{*}{\makecell[c]{test-\\clean}} & \multirow{2}{*}{\makecell[c]{test-\\other}} & \multirow{2}{*}{tt} & \multirow{2}{*}{dev} & \multirow{2}{*}{\makecell[c]{test-\\clean}} & \multirow{2}{*}{\makecell[c]{test-\\other}} \\
  	  & small & medium & large &  &  &  &  &  &  &  &  \\
	\midrule
  	1 & \multicolumn{3}{c|}{-} & \checkmark & 61.4 & 85.8 &  86.5 & 87.4 & 99.20 & 45.46 & 44.05 & 46.49  \\
  	\midrule
  	2 & \checkmark &  &  & - & 76.1 & 57.8 & 59.5 & 61.0 & 70.60 & 45.19 & 42.56 & 46.84 \\
  	3 &  & \checkmark &  & - & 43.9 & 27.2 & 25.7 & 27.5 & 54.80 & 54.12 & 52.89 & 56.55 \\
  	4 &  &  & \checkmark & - & 28.1 & 21.0 & 19.0 & 21.7 & 77.30 & 55.48 & 54.18 & 56.25 \\
  	\midrule
  	5 & \checkmark &  &  & \checkmark & 23.9 & 45.5 & 45.5 & 49.4 & 99.30 & 50.80 & 51.12 & 53.94  \\
  	6 &  & \checkmark &  & \checkmark & 15.1 & 23.6 & 22.9 & 24.5 & 99.70 & 55.41 & 55.25 & 58.31 \\
   	7 &  &  & \checkmark & \checkmark & 13.6 & 21.0 & 19.6 & 21.4 & 99.40 & 59.73 & 58.14 & 59.79  \\
  	\bottomrule
  	\bottomrule
  \end{tabular}
  }
  \vspace{-10pt}
\end{table}

The WHAMR! \cite{Maciejewski2020WHAMR} dataset is a public benchmark built upon wsj0-2mix \cite{hershey2016deep} and WHAM! \cite{Wichern2019WHAM} for the task of overlapped speech separation and recognition under reverberant and noisy conditions. 
It serves as one of the publicly available benchmarks for speech separation and recognition under reverberant and overlapping conditions. 
Further experiments were conducted to evaluate the generalizability of the proposed LibriheavyMix dataset. 
Note that all experiments involving WHAMR! use the {\it clean\_reverb} data to match the acoustic environment of LibriheavyMix. 

The performance of Conv-TasNet models is presented in Tab. \ref{tab:whamr_enh}.
All models were trained on {\it min} version of the WHAMR! and LibriheavyMix and evaluated on {\it max} test sets of the corresponding corpora.
Sys. 1 suggests that the model trained on WHAMR! performs not as well on LibriheavyMix as it demonstrates on WHAMR!.
This observation aligns with the previous study \cite{kadiouglu2020empirical} indicating that the Conv-TasNet model trained on wsj0-2mix demonstrates poor generalization on other datasets.
Sys. 2-4 suggest that by introducing more diversity into the training data and scaling up the amount of training data, the Conv-TasNet model achieved a significant improvement in generalization even on the unseen WHAMR! data. 
The performance on both LibriheavyMix and WHAMR! can be further boosted when incorporating training data from WHAMR! dataset, as Sys. 5-7 consistently outperform Sys. 2-4  on all test sets involved.

The performance of the SOT models is presented in Tab. \ref{tab:whamr_asr}.
Results suggest that scaling the amount of training data demonstrates a significant WER reduction especially on the most complicated 4-speaker scenario. 
Although a similar trend is also observed in terms of the speaker counting accuracy, it is still challenging especially when multi turn conversations are presented in test sets.

%\subsection{Cascaded Conv-TasNet + SOT Pipelined System}
%
%\input{tab/cascaded}
%
%To further investigate the impact of the amount of training data on the speech recognition system, a cascaded system is constructed on top of trained Conv-TasNet and SOT systems and further evaluated on the WHAMR! and proposed dataset.

\subsection{Performance of Speaker Diarization Baseline}

\begin{table}[t]
  \caption{Performance of the pyannote.audio diarization system and cascaded systems on LibriheavyMix test sets.}
  \label{tab:diar}
  \vspace{-8pt}
  \centering
  \resizebox{\linewidth}{!}{
  \begin{tabular}{c|c|c|c|c}
  	\toprule
  	\toprule
	\multirow{2}{*}{Sys.} & \multirow{2}{*}{\makecell[c]{Test Set \# spkr}} & \multicolumn{3}{c}{Diarization Error Rate (\%) $\downarrow$}    \\
	&  & dev & test-clean & test-other \\
	\midrule
	\multirow{3}{*}{1} & 2 & 30.90 & 31.72 & 28.20 \\
	& 3 & 42.13 & 41.96 & 40.17 \\
	& 4 & 50.27 & 48.49 & 47.42 \\
	\midrule
	2 (Sys. 4, Tab. \ref{tab:whamr_enh} $\rightarrow$ Sys. 1, Tab. \ref{tab:diar}) & 2 & 19.68 & 21.20 & 19.47 \\
	3 (Sys. 7, Tab. \ref{tab:whamr_enh} $\rightarrow$ Sys. 1, Tab. \ref{tab:diar}) & 2 & 19.39 & 21.03 & 19.40 \\
	\midrule
	\midrule
	\multirow{2}{*}{Sys.} & \multirow{2}{*}{\makecell[c]{Test Set \# spkr}} & \multicolumn{3}{c}{Word Error Rate (\%) $\downarrow$} \\
	&  & dev & test-clean & test-other \\
	\midrule
	4 (Sys. 4, Tab. \ref{tab:whamr_enh} $\rightarrow$ Sys. 4, Tab. \ref{tab:whamr_asr}) & 2 & 44.9 & 42.5 & 47.1\\
	5 (Sys. 7, Tab. \ref{tab:whamr_enh} $\rightarrow$ Sys. 7, Tab. \ref{tab:whamr_asr}) & 2 & 43.4 & 41.0 & 45.9\\
  	\bottomrule
  	\bottomrule
  \end{tabular}
  }
  \vspace{-15pt}
\end{table}

For simplicity, speaker diarization was directly performed on the dev and test sets of LibriheavyMix using pre-trained pyannote.audio 3.1 system. 
Performance of the pyannote.audio system is presented in Tab. \ref{tab:diar}.
The speech separation module demonstrates its effectiveness by delivering a remarkable absolute DER reduction of up to 11.51\% to the diarization system. 

\section{Conclusions}
This work releases a large-scale (20,000 hours) synthesized corpus \footnote{\url{https://huggingface.co/datasets/zrjin/LibriheavyMix-{dev,test,small,medium,large}}}. for overlapped speech separation and recognition under reverberant conditions. 
A series of baseline systems are constructed to evaluate the performance of the proposed dataset. 
Further evaluation using a public benchmark for far-field overlapped speech separation and recognition validates the effectiveness and generalizability of the proposed dataset.

\newpage

\bibliographystyle{IEEEtran}
\bibliography{mybib,bib/enh,bib/mtk_asr,bib/datasets,bib/rir,bib/1intro/asr,bib/3simu/simu,bib/diar}

\end{document}